# Field electron emission mechanism in an ultrathin multilayer planar cold cathode

Ru-Zhi Wang[1*]   Hui Yan[1*]   Bo Wang[1]   Xing-Wang Zhang[2]   Xiao-Yuan Hou[3]

[1] Laboratory of Thin Film Materials, Beijing University of Technology, Beijing 100022, China

[2] Key Laboratory of Semiconductor Materials Science, Institute of Semiconductors, Chinese Academy of Sciences, P. O. Box 912, Beijing 100083, China

[3] Surface Physics Laboratory (National Key Laboratory), Fudan University, Shanghai 200433, China

Field electron emission from an ultrathin multilayer planar cold cathode (UMPC) including quantum well structure has been both experimentally and theoretically investigated. We found that only tuning the energy levels of UMPC the field electron emission (FE) characteristic can be evidently improved, which is unexplained by the conventional FE mechanism. Field electron emission mechanism dependent on the quantum structure effect, which supplies a favorable location of electron emission and enhances tunneling ability, has been presented to expound the notable amelioration. An approximate formula brought forward can predict the quantum FE enhancement, which the theoretical prediction is close to the experimental result.

## I. Introduction

The cold cathodes have been attracted more and more attentions for its important applications in flat-panel display and some power amplifier [1]. However, the present cold cathodes do not have so good performance that it can be considered as a commercial application on a large scale. As an effective conventional technique, the nanoscale protrusion (or microtip) cold cathodes have been widely used to lower the threshold voltage enabling field emission (FE) by utilizing the geometric field enhancement effect; however, its fabrication processes are so complicated that make its price too expensive. As for the heartening carbon nanotubes (CNTs), a too high FE current is unstable and may lead to a vacuum breakdown [2]. People tend to consider the planar cold cathodes constructed by some wide band-gap semiconductor (WBGS) films [3], which owes to its simple fabrication, easy integration, convenient control, and stable emission et al. Nevertheless, its FE current density is not high enough and/or the operational voltage is not low enough [1]. Therefore, best of all, it is necessary to advance electron emission properties of the planar cold cathodes for potential device. For the planar cold cathode, there are presently three feasible mechanisms except for the geometric field enhancement to advance its electron emission properties. One is the Schottky diode with a negative-electron-affinity (NEA) semiconductor surface [4]; the second is the composite of the electric-field enhancement, and Schottky diode with a NEA semiconductor [5]; the last is the surface barrier lowered with an ultrathin wide band-gap semiconductor layer (UTSC) [6]. For these mechanisms, their essential goal is to make electrons more easily tunnel by the reduction of the surface potential barrier. Most recently, it is interestingly found that the quantum well states and surface resonance states in ultrathin films can be characteristic by the field electron emission (FE) current [7, 8]. The previous theoretical analysis [9-15] also showed that, due to electron confinement in quantum well, electron emission from the ultra-thin cold cathodes presents a distinct resonant behavior. Since the tunneling ability of electrons corresponding to resonant states in ultrathin films can be evidently affected by the quantum structure [10], it impliedly suggests an electron emission mechanism to improve FE characteristic by modulating quantum structure of ultrathin films. In this paper, we demonstrate the field electron emission mechanism by examining FE from an ultralthin 6nmGaAs/3nmAlAs or 3nmGaAs/6nmAlAs two-layer planar cold cathode. We found that by only modulating the quantum structure the FE current is evidently enhanced, and the threshold voltage is also reduced, which can not be explained by the previous FE mechanism. Ulteriorly theoretical analysis based on the quantum self-consistent scheme [14] show the amelioration of electron emission from ultra-thin multilayer planar cold cathodes (UMPC) is resulted from a favorable emission location and an advantaged electron tunneling due to the quantum structure effect. Furthermore, we present an approximate formula to predict the quantum FE enhancement.

___________
*To whom correspondence should be addressed: wrz@bjut.edu.cn; hyan@bjut.edu.cn

## II. Results and discussion

The inset of figure 1 shows the schematic drawing of the planar cold cathode structure. The cathode structure was grown by an EPI Gen-II solid-source molecular beam epitaxy (MBE) system on an $n^+$ type GaAs (001) substrate. After the native oxide was desorbed at 580 $^o$C under As atmosphere, the substrate was heated up to 600 $^o$C. A 170nm-thick Si-doped GaAs buffer layer ($n=1\times10^{18}$ cm$^{-3}$) was grown initially, followed by the GaAs/AlAs two-layer planar cold cathode. To investigate the effect of the quantum structure of the planar cathodes on the electron emission, two samples, 3nm GaAs/ 6nmAlAs for sample A and 6nm GaAs/3nm AlAsfor sample B, were prepared. For both samples, the growth temperature is 600 $^o$C and the V/III beam equivalent pressure (BEP) ratio used is 20. The growth rate of GaAs and AlAs is 1ML/s. The surface of both samples was characterized to be atomically smooth by atomic force microscope (AFM) over the whole surface of the cathode. Electron field emission experiments were performed in an ultrahigh vacuum chamber with a base pressure better than $4\times10^{-9}$ Torr. The current density-field characteristics were measured at room temperature using a diode structure with a low resistivity (0.02 Ω cm) silicon wafer as the anode. The sample and the silicon were separated by two pieces of round glass fibre as a spacer. The typical diameter of the round glass fiber was 14μm measured by a precision gauge with an accuracy of ±0.5μm.

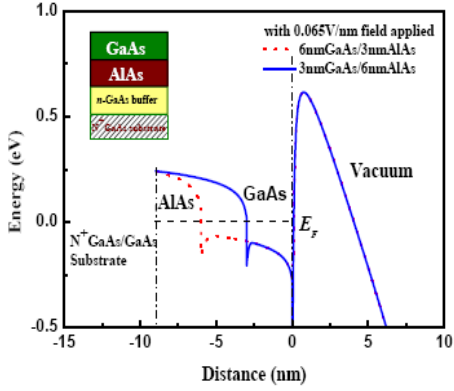

Fig.1 (Color online) Calculated energy band diagram of two UMPCs structure with the field of 0.065V/nm applied

As shown in Fig.2, the current densities versus electric filed (J-E) curves are evaluated to show field emission characteristics for two UMPC samples with different quantum structures (see Fig.1). A obvious distinction is demonstrated in J-E curves which can be more clearly found in the Fowler-Nordheim (F-N) plot (see the inset of Fig. 2) that the current densities are enhanced to about 16 times by the quantum structure effect. In addition, if the threshold voltage is defined at an emission current density of 0.1μA/cm$^2$, the threshold voltage decreases from 56V/μm to 43V/μm. These results indicate the novel effect of quantum structure on FE characteristic of the UMPC. For an ideal semiconductor film with atomically smooth surface, if considered the potential drop (barrier field) χ is linearly determined by the field density, then $\chi = \gamma F$, the F-N equation can be simplified as [16]:

$$\ln(J/F^2) = \ln\left(\frac{A\gamma^2}{\phi}\right) - \frac{B\phi^{3/2}}{\gamma F} \quad (1)$$

where J is current density, χ is the potential drop, $\phi$ is the height of surface potential barrier, and A and B are constants, respectively. Here, the potential distribution (Fig. 1) can be calculated by the self-consistent scheme of solving the Poisson's equation [14], and the relative experimental parameters are selected from the handbook [17] in these calculations. The surface potential barriers ϕ from the calculated energy band in Fig.1 have the same shape and height when the same field applied to two UMPCs. It means that, for an electron in GaAs quantum well with the identical incident energy, there will be the same emission ability from the surface barrier. From the inset of Fig.2, it can be easily found that the slope $\frac{B\phi^{3/2}}{\gamma}$ of the two F-N plots is almost identical. These results naturally lead to an equal λ and then an exactly same theoretical FN curve for two UMPCs by Eq. (1). Obviously, the theoretical analysis will be very conflictive with the experiment measurement in Fig.2. Therefore, electron emission from a UMPC can not be explained by the conventional field emission mechanism in which the reduction of the surface potential barrier is considered as the most important factor to improve field emission from semiconductor surfaces [4-6] or an UMPC. The present experiment may suggest field electron emission mechanism being independent of the surface potential barrier. What is the origination of the remarkably enhanced field emission current? There may be some useful indexes [7-15] to help understand the effect of the quantum structure for the UMPC. However, all previous reports, even in our theoretical investigation [14] assumed the enhancement on field emission current should be mainly benefited from the surface barrier lowered. It is necessary to explore the essence of electron emission by analyzing the effect of quantum structure in an UMPC.

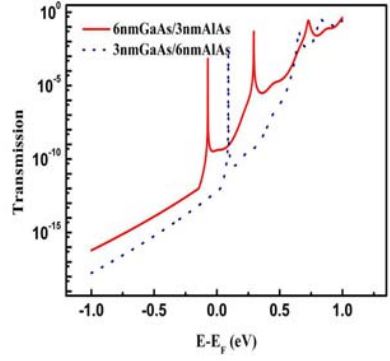

Fig.3 (Color online) Resonant transmission in two UMPCs structures with the field of 0.065 V/nm applied.

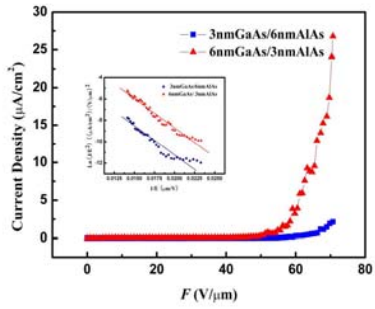

. Fig.2 (Color online) Electron emission current in two UMPCs structures (the inset is the corresponding F-N plots).

As shown in Fig.1, the band structure have been calculated out by the self-consistent scheme including the band bending and the more realistic image potential [14], and it is obvious that there are very different for the energy band structure of two UMPCs. Since the emitter band structure consumedly affect the field electron energy distribution (FEED) [18], therefore, to explore the essence of field emission from UMPC, it will be a useful index to analysis its FEED. In general, the FE current from a semiconductor film can be written as [15]:

$$J = \frac{4\pi q m_t k_B T}{h^3} \int T(E_x) \ln[1 + e^{-(E_x - E_F)/k_B T}] dE_x = J_0 \int J(E_x) dE_x = J_0 J_T$$

(2)

and

$$J(E_x) = T(E_x) \ln[1 + e^{-(E_x - E_F)/k_B T}] = T(E_x)\lambda(E_x)$$

(3)

where $J_0 = 4\pi q m_t k_B T / h^3$, which is mainly determined by the electron effective transverse mass $m_t$, $J_T$ is defined by the tunneling factor of the FE structure, which is the integral of the field emission energy distribution (FEED) $J(E_x)$, $q$ unit charge, $k_B$ Boltzmann's constant, $T$ the temperature, $h$ Plank's constant, and $E_F$ Fermi energy. $T(E_x)$ is the tunneling probability of the FE structure, which can be obtained by the quantum transfer matrix (TM) method [19,20] since the potential distribution of the UMPC can be addressed by the quantum self-consistent scheme [14]. It is clearly seen that the conduction band minimum (CBM) of GaAs layer is below the Fermi energy even to 0.25eV for forming a quantum well. It should owe to the strong band bending in ultrathin wide bandgap semiconductors films [21]. The band structure characteristics are also presented by the calculated $T(E)$ with the electron incident energy in Fig. 3, it is obvious that there are three distinct peaks when electrons tunnel the UMPC structure. There are two quantum energy levels localized in the GaAs quantum well for 6nmGaAs/3nmAlAs, however, only one quantum energy level appears in the GaAs quantum well for 3nmGaAs/6nmAlAs. The emission current may be dependent on a favorable electron emission location. Now let's estimate quantitatively the effect of the quantum structure on FE enhancement. As it is well known, the effect of the quantum structure on electron emission location can be reflected by the change of the effective mass. From Eq. (2), $J_0$ is decided by the effective transverse mass $m_t$. Considered the AlAs/GaAs as an integrated structure, we adopted the weighted averages of the effective mass [22] as the calculated $m_t$, which may be expressed by linearly combined with the thickness ratio of GaAs and AlAs.

$$m_t = r_1 m_{GaAs} + r_2 m_{AlAs} \qquad (4)$$

where $m_{GaAs}$, $m_{AlAs}$ and $r_1$, $r_2$ are the effective transverse mass and thickness ratio of GaAs and AlAs, respectively. The FE current of 6nmGaAs/3nmAlAs calculated from Eq. (4) is about 0.718 times than that of 3nmGaAs/6nmAlAs. In the calculations $m_{GaAs}$ is $0.063m_0$ and $m_{AlAs}$ is $0.19m_0$ ($m_0$ is electron unit mass). However, the experimental current densities are enhanced to about 16 times, and it may indicate another notable FE enhancement by the quantum structure effect.

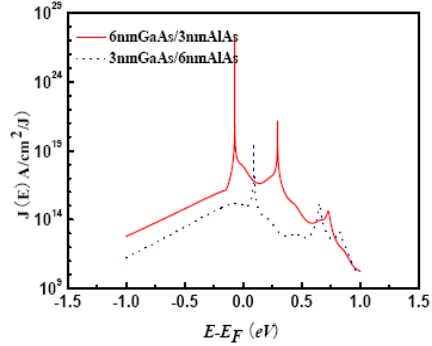

Fig.4 (Color online) FEEDs of two UMPCs structures.

To further demonstrate the essence of electron emission from the modulated quantum structure, the FEEDs of two UMPCs are also presented in Fig. 4, and theirs curves hold some resonant peak characteristics. This is also approved in the calculated FEEDs of Ultrathin Films of Fe on W (001) by the density functional calculations (DFT) theory [8]. In Fig. 4, the curve of the FEEDs is raised by orders of magnitude due to the different modulation of the quantum structures in the two UMPCs. Because of optimizing the quantum structure compared with the 3nmGaAs/6nmAlAs, the tunneling ability of all electrons in 6nmGaAs/3nmAlAs is consumedly upgraded. Naturally, it will advance the tunneling factor $J_T$ to enhance the FE current from Eq. (2). Since the variety in FE current by modulating $m_t$ is about 0.718 times, it means that the tunneling factor exaltation $J_T$ is a more effective mechanism to improve the FE characteristic for the present UMPCs. To advance the $J_T$, a resultful scheme is to modulate the quantum energy level toward low energy section according to Eq. (3). From Fig. 4, the current density is mainly supplied from these electrons around the resonant energy level. However, due to the band bending and the image potential, the band structure is asymmetric in Fig. 1, therefore, field emission tunneling will be more complicated, and the width of the resonant state of field emission depends on the barrier heights and widths as well as the energy level [23-26]. Considering near a sharp (resonance) maximum, $T(E_x)$ can be simply written as a Lorentzian form [27]:

$$T(E_x) = \frac{T_i}{1 + ((E_x - E_i)/\Delta E_i)^2} \quad (5)$$

Where $E_i$ is the ith quasi level energy of UMPCs, $T_i$ is the maxium transmission probability of the ith quasi level energy, and $\Delta E_i$ is the resonance half-width of the ith quasi level energy, respectively. To extrude the effect of the resonant transmission probability, the computed energy range of T(E) is considered to locate in $2\Delta E$. Additionally, there is only a tiny shift of the position of resonant energy level with the different field applied [14]. Then, we can approximately predict the relative FE enhancement of two UMPCs by quantum resonant field emission by the following formula:

$$\beta_q = \frac{m_{t1}}{m_{t2}} \left\{ \frac{\sum_{i1} \int T(E_x) \lambda(E_x) dE_x}{\sum_{i2} \int T(E_x) \lambda(E_x) dE_x} \right\}$$

$$= \frac{m_{t1}}{m_{t2}} \left\{ \frac{\sum_{i1} \int_{|E_x - E_{i1}| \leq 2\Delta E_{i1}} \frac{T_{i1}}{1 + ((E_x - E_{i1})/\Delta E_{i1})^2} \ln(1 + e^{-(E_x - E_F)/k_B T}) dE_x}{\sum_{i2} \int_{|E_x - E_{i2}| \leq 2\Delta E_{i2}} \frac{T_{i2}}{1 + ((E_x - E_{i2})/\Delta E_{i2})^2} \ln(1 + e^{-(E_x - E_F)/k_B T}) dE_x} \right\}$$

(6)

where $m_{t1}$, $m_{t2}$ and $E_{i1}$, $E_{i2}$ are the effective transverse mass and the resonant energy level of two UMPC, respectively, and $\beta_q$ is defined as the quantum structure relative enhancement factor. It may be understood for the optimum quantum structure that, it not only supplies the most favorable emission location by the energy level shift but also produces the best tunneling ability by electron accumulation in quantum well. If $|E_i - E_F| \gg k_B T$, then $\lambda(E_x)$ will tend to zero when $\Delta E$ is very narrow, the contribution to $\beta_q$ of the ith energy level may be omitted. For the present two UMPCs, T=300k and $k_B T$ is about 0.026eV, therefore, as shown in Fig.3, the effect of $\beta_q$ mainly resulted from only one quasi energy level of every UMPC, respectively. The calculated $\beta_q$ from Eq. (6) is about 9.16, which may be comparative value with the experimental FE enhancement time of about 16. For the experimental value is larger than theoretical prediction, it may be originated from the discrepancy of numerical approximation and other effect of field enhancement in the UMPCs, which can be explained from the surface potential reduction by the previous FE mechanism.

## III. Conclusion

In summary, we have demonstrated field electron emission mechanism by both experimentally and theoretically analyzing the FE proprieties of the ultrathin multilayer planar cold cathodes. Not as well as the previous field electron emission mechanism of cold cathodes, the elevation of the FE current or the reduction of the threshold voltage is independent of the depression of the surface potential barrier or the enhancement of the field density. FE current can be remarkably enhanced by only modulating the quantum structure of the cathodes. And we also bring forward a formula to approximately predict the quantum FE enhancement. Since the modulation of the quantum structure by a simple technique is easier than the control of the surface potential barrier by a complicated method for ultrathin films, it is more possible to design an excellent nanostructured cold cathodes device by the present mechanism.

**Acknowledgements**

This work is supported by PHR (IHLB), the National Natural Science Foundation of China (No.10604001, No.60576012), and the Natural Science Foundation of Beijing (No. 4073029). Experimental assistance and fruitful discussions by Prof. W. Q. Ma from Institute of Semiconductors, CAS, is gratefully acknowledged.